\documentclass{ws-p8-50x6-00}

\begin{document}


\thispagestyle{empty}

\begin{flushright}
IRB-TH-10/99\\
December, 1999
\end{flushright}

\vspace{2.0cm}

\begin{center}
\Large\bf Preasymptotic effects in beauty decays
\vspace*{0.3truecm}
\end{center}

\vspace{1.8cm}

\begin{center}
\large B. Guberina, B. Meli\'{c} and
 H. \v Stefan\v ci\'c\\
{\sl Theoretical Physics Division, Rudjer Bo\v skovi\'c Institute,
\\
P.O.Box 1016, HR-10001 Zagreb, Croatia\\[3pt]
E-mails: {\tt guberina@thphys.irb.hr,
         \tt melic@thphys.irb.hr,
         \tt shrvoje@thphys.irb.hr}}
\end{center}

\vspace{1.5cm}

\begin{center}
{\bf Abstract}\\[0.3cm]
\parbox{13cm}
{
Large preasymptotic effects in beauty decays have been found
using heavy-quark and SU(3) symmetry, as well as experimental
data on charmed hyperons.
Contrary to rather uniform beauty-meson lifetimes, a much larger
spread of beauty-baryon lifetimes is predicted. However,
it is highly unlikely that, theoretically, the
$ \tau(\Lambda_b)/\tau(B_d^0) $ ratio, which
at present deviates more than $1\sigma$ from the experimental result,
can be lowered below 0.9.
}
\end{center}

\vspace{2.5cm}

\begin{center}
{\sl Talk given by B. Meli\'c at 
The $3^{\rm rd}$ International Conference on B Physics and CP Violation, 
Taipei, Taiwan, December 3-7, 1999\\
To appear in the Proceedings (without figures)}
\end{center}

\vbox{}
\thispagestyle{empty}
\newpage
\setcounter{page}{1}
\thispagestyle{empty}


\title{Preasymptotic effects in 
beauty decays\footnote{Talk given by B. Meli\'c at
The $3^{\rm rd}$ International Conference on B Physics and CP Violation,
Taipei, Taiwan, December 3-7, 1999}}

\author{B. Guberina, B. Meli\'c, and H. \v Stefan\v ci\'c}

\address{
Theoretical Physics Division, 
Rudjer Bo\v{s}kovi\'{c} Institute, 
P.O.Box 1016, \\HR-10001 Zagreb, Croatia
\\E-mails: guberina@thphys.irb.hr, melic@thphys.irb.hr, shrvoje@thphys.irb.hr}


\maketitle

\abstracts{
Large preasymptotic effects in beauty decays have been found 
using heavy-quark and SU(3) symmetry, as well as experimental 
data on charmed hyperons.
Contrary to rather uniform beauty-meson lifetimes, a much larger 
spread of beauty-baryon lifetimes is predicted. However, 
it is highly unlikely that, theoretically, the 
$ \tau(\Lambda_b)/\tau(B_d^0) $ ratio, which 
at present deviates more than $1\sigma$ from the experimental result, 
can be lowered below 0.9.
}

The saga of beauty never comes to an end\cite{bigi}. 
During the last decade the general belief 
has been that decays of beauty quarks should be 
very well described theoretically in the framework of the 
Operator Product Expansion 
(OPE) and the Heavy-Quark Effective Theory (HQET). The mass of the 
beauty quark, being of the order of 5 GeV, appears to be heavy enough 
to ensure the fast convergence of the $1/m_b$ expansion. The diversity 
of lifetimes of 'beautiful' mesons and baryons is expected first 
at the subleading level in the $1/m_b$ expansion. The lifetimes of beauty 
mesons follow this simple theoretical prediction within $5 - 10\%$ 
in the $m_b \rightarrow \infty$ limit:
\begin{equation}
\tau(B^{+})=\tau(B_{d}^{0})=\tau(B_{s}^{0}) \,.
\end{equation}
The only measured baryon lifetime $\tau(\Lambda_b)$ appears to be 
smaller by $15-25 \%$; experimentally, it follows 
that
\begin{equation}
\frac{\tau(\Lambda_{b})}{\tau(B_{d}^{0})} = 0.81 \pm 0.05 \; 
{\rm (PDG)}\, ,
\label{ratio}
\end{equation}
while the theoretical value is about 0.98. This discrepancy 
raises doubt 
that the quark-hadron duality might be, {\it horribile dictu}, severely flawed. 

\begin{figure}
\centerline{\epsfig{file=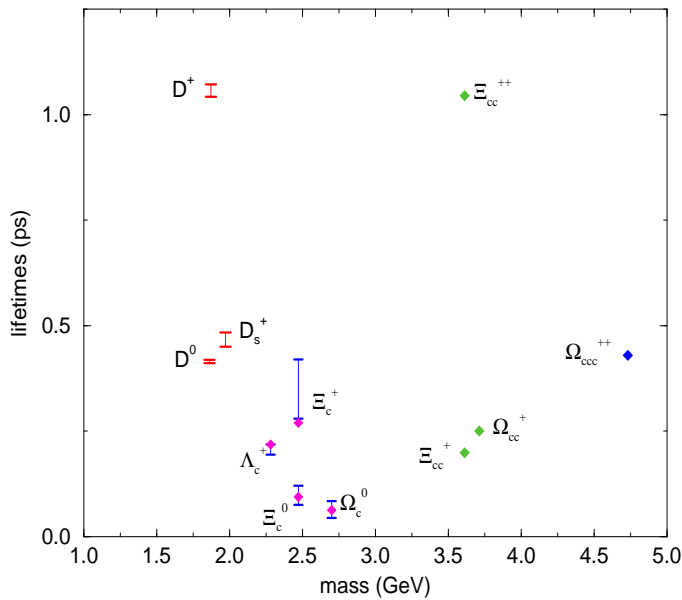,height=8cm,width=9cm,silent=}}
\caption{Experimental (with error bars) and theoretical (filled diamonds - 
calculated in {\protect\cite{GM}} and {\protect\cite{GMS}}) results 
for lifetimes of weakly decaying charmed hadrons.}
\label{f:fig1}
\end{figure}

The rate of the beauty-hadron decay is given as a sum over matrix 
elements of D-dimensional operators. 
The sum starts with the operator of dimension D=3 and shows the 
fast convergence in the $1/m_b$ expansion. 
First corrections proportional to the operator of D=5 are controllable 
and by taking them into account 
we obtain the difference of $2-3\%$ in the lifetimes of beauty hadrons.
Therefore, the only hope to come closer to the ratio (\ref{ratio}) 
is to look for the possible larger contributions coming from 
the operators of dimension D=6 or higher. 
These operators are 
known to play an important role in charmed-meson decays\cite{mesons} and 
their effects are 
even more pronounced in charmed-baryon decays\cite{baryons}. 
Recently, the analysis of singly charmed 
baryons\cite{GM} has been extended to doubly 
charmed baryons\cite{GMS,Rusi} (Fig.1), where the 
idea  of 'meson-like' baryons, in which heavy quarks form a diquark, 
was applied. A possible 
physical interpretation of this philosophy is the existence of hadronic 
supersymmetry\cite{Lich}, which appears whenever the diquark is physically 
realizable. 

However, the calculation of the contribution of D=6 operators suffers 
from the strong model-dependence in the evaluation of their matrix elements.
Recently, Voloshin\cite{Vol} has proposed the way of reducing the model 
dependence. His main assumption is that, owing to SU(3)-flavor 
and heavy-quark symmetry, the matrix elements of four-quark operators 
extracted from experimental data on charmed hyperons can be used for 
calculation in the beauty sector, provided that they are normalized at 
the low energy scale $\mu \sim 1\;{\rm GeV}$. The result of these assumptions 
is that lifetime differences between beauty hyperons $(\Lambda_b, \Xi_b)$ can 
be expressed through the (measured) lifetime differences between charmed 
hyperons $(\Lambda_c, \Xi_c)$ without invoking an explicit model-dependent 
calculation of four-quark operators. The possible uncertainty of this 
approach is of the order ${\cal O}(m_c^2/m_b^2) \sim 0.1$. 

We have extended\cite{mi} Voloshin's analysis by introducing a rather 
modest model dependence in order to obtain more predictive power, 
i.e., the lifetimes 
of the $(\Lambda_b, \Xi_b)$ hyperon triplet and the lifetime of $\Omega_b$. 
Basically, we express the decay rates in terms of the baryon wave function 
squared $|\Psi(0)|^2$, which parametrizes the four-quark operator 
contributions and is usually given by the 
nonrelativistic relation $|\Psi(0)|^2 \sim f_B^2$, $f_B$ being the meson decay 
constant. We use Voloshin's approach to determine the value 
of the $|\Psi(0)|^2$. From the lifetime 
differences between members of beauty hyperons we are able to 
extract an effective decay constant, $|\Psi(0)|^2 \sim (F_B^{eff})^2$, 
\begin{equation}
F_{B}^{eff} = (0.441 \pm 0.026) \, {\rm GeV} \, .
\end{equation}
It is instructive to compare this value with the value of $f_B$, 
$f_B = 0.16-0.17 \,{\rm GeV}$, and then the ratio of the baryon over meson 
wave function, which is, according to our model, about 7-8 times larger 
than when we apply the nonrelativistic model:
$|\Psi_{\Lambda_b}(0)|^2/|\Psi_{B}(0)|^2 \sim 4.2$. 
This effectively means that, in our approach, four-quark contributions 
in beauty-baryon lifetimes are enhanced by almost an order of 
magnitude.

Owing to ambiguities in the determination of the bottom quark mass, we 
concentrate mainly on the lifetime ratios. We follow the 
philosophy of using the running quark mass ${\overline m}_b$ to avoid 
the renormalon ambiguities and therefore have to choose 
${\overline m}_b(1\, {\rm GeV}) = 4.7 \, {\rm GeV}$.

The predicted hierarchy in the sector of beauty baryons is then
\begin{equation}
\label{eq:hierarchy}
\tau(\Lambda_{b}) \simeq \tau(\Xi_{b}^{0}) < \tau(\Xi_{b}^{-}) <
\tau(\Omega_{b}) \, ,
\end{equation}
and the obtained  lifetime ratios are 
\begin{equation}
\tau(\Xi_b^-)/\tau(\Lambda_b) \simeq 1.20\, ,  \qquad\qquad 
\tau(\Omega_b)/\tau(\Lambda_b) \simeq 1.30 \, .
\end{equation}
These ratios are much larger than those 
predicted by the standard nonrelativistic model.

Let us now briefly discuss the problem of the $\tau(\Lambda_b)$ over 
$\tau(B)$ ratio. 
The problem is the clear discrepancy between the experimental ratio 
(\ref{ratio}) and the theory which gives 
\begin{equation}
\tau(\Lambda_b)/\tau(B_d^0) = 0.97 + {\cal O}(1/m_b^3)
\end{equation}
just by taking the first, ${\cal O}(1/m_b^2)$ corrections into account. 
Owing to the fast convergence of the $1/m_b$ expansion and because 
the vacuum saturation approximation for 
mesons works rather well, it seems that the decay rate of 
the $B$-meson cannot be significantly smaller to lower 
the $\tau(\Lambda_b)/\tau(B)$ 
ratio. The only hope then persists in the enlargement of four-quark 
contributions in the $\Lambda_b$-decay, but these effects also 
cannot be pushed over some limit. 
So the question is: can we accommodate the theoretical prediction 
on the $\tau(\Lambda_b)/\tau(B_d^0)$ ratio to the experimental 
result, using our enhancement of the four-quark 
contributions in the $\Lambda_b$-decay ?

Clearly, we need the smaller $m_b$ to obtain the larger preasymptotic 
effects, but there is a competition between the ${\cal O}(1/m_b^2)$ 
effects in mesons and the ${\cal O}(1/m_b^3)$ effects in $\Lambda_b$. 
The net result is then the stable $\tau(\Lambda_b)/\tau(B)$ ratio: 
\begin{equation}
\tau(\Lambda_b)/\tau(B_d^0) \sim 0.90 \pm 0.01 \qquad {\rm for} \qquad 
m_b = 4.4 - 4.8 \, {\rm GeV}\, .
\label{ratioTH}
\end{equation}

We have also checked the result against the deviation from 
the valence quark approximation (VQA)\cite{Vol2}, which 
equals the 
deviation of the ${\overline B}$-parameter from one:
\begin{equation}
x = - {\overline B} y \, .
\end{equation}
Here $ x \sim \langle {\overline b} \Gamma_{\mu} b\, 
{\overline q} \Gamma^{\mu} q \rangle$ and $ y \sim \langle {\overline b}^i 
\Gamma_{\mu} b^j\, 
{\overline q}^j \Gamma^{\mu} q^i \rangle$, $\Gamma_{\mu} = (V - A)_{\mu}$, 
 and $q$'s denote light quarks in a baryon. 

We have extracted this ratio to be 
$|x/y| \approx 1.8 \pm 1.0$ at $\mu = 1\, {\rm GeV}$, which is 
consistent with the result\cite{Vol2} of Voloshin. Owing to the large  
error in its determination, it is difficult to give a 
definite conclusion about the validity of the VQA and therefore we prefer 
to use ${\overline B} = 1$ in our predictions. 
However, it is also clear 
that the VQA cannot be generally valid, because of the fact that $y$ is 
a $\mu$ dependent quantity, while $x$ is not, 
and the result that $|x/y|$ is significantly larger than 1 
might prove at the end to be both correct and fundamental.
But, even if the VQA is heavily broken by almost $100 \%$, 
the $\tau(\Lambda_b)/\tau(B)$ ratio {\it cannot} be lowered below 0.9 value.

\begin{figure}
\centerline{\epsfig{file=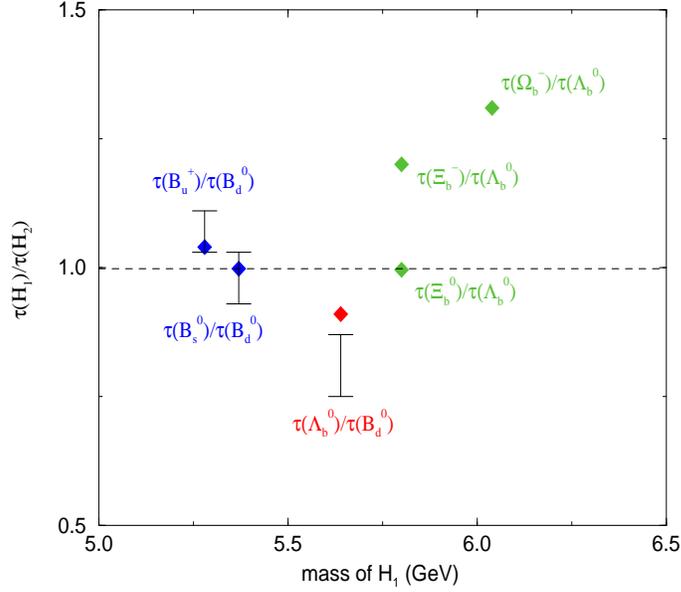,height=8cm,width=9cm,silent=}}
\caption{Experimental (with error bars) and theoretical (filled diamonds) 
results for ratios of lifetimes of beauty hadrons. Meson lifetimes are 
obtained using $f_b = 160 \, {\rm GeV}$.}
\label{f:fig2}
\end{figure}

To conclude, although there might still be some place for 
nonfactorizable contributions in mesons to play 
some role, it is highly unlikely that 
the lifetimes of $\Lambda_b$ and the $B$-meson
can be split by more than $10\%$ (\ref{ratioTH}). To 
reach the experimental value for the $\tau(\Lambda_b)/\tau(B)$ ratio 
(\ref{ratio}) would require $F_B^{eff}|_{\rm fit} \sim 0.720 \,{\rm GeV}$, 
which can hardly be accommodated in the present theory. 
Should future data maintain the $\tau(\Lambda_b)/\tau(B)$ ratio 
well below 0.9, that would indicate the violation of some of underlying 
concepts of the present theory, such as the quark-hadron 
duality. 
One of the tests will also be the experimental check 
of the predicted spread of beauty-baryon lifetimes of the order 
of $20 \%$ in the $\tau(\Xi_b^-)/\tau(\Lambda_b)$ ratio and 
of $30 \%$ in the $\tau(\Omega_b)/\tau(\Lambda_b)$ ratio (Fig.2).

\section*{Acknowledgments}
B.M. would like to thank 
H.Y. Cheng and W.S. Hou for invitation to participate in this 
very stimulating conference.
This work was supported by the Ministry of
Science and Technology of the Republic of Croatia under Contract No.
00980102.

\end{document}